\documentclass[usenatbib,conference]{basi}
\usepackage[british]{babel}
\usepackage[varg]{txfonts}
\usepackage{rotating}
\usepackage{dcolumn}
\begin{document}
\title[ Stellar Parametrization]{ The Stellar parametrization  using Artificial Neural Network  }

 \author[Sunetra Giridhar et al. ]%
         {Sunetra Giridhar$^{1}$ \thanks{e-mail: giridhar@iiap.res.in;
    aruna@iiap.res.in;  akunder@ctio.noao.edu, muneers@iiap.res.in; selva@iiap.res.in}
     Aruna Goswami$^{1}$, Andrea Kunder $^2$, S. Muneer $^3$, G. Selva Kumar $^4$\\ 
    $^1$  Indian  Institute of Astrophysics, Bangalore 560 034, India \\
    $^2$  Cerro Tololo Inter-American Observatory, NOAO, Casilla 603, La Serena, Chile \\
    $^3$  CREST Campus, Indian  Institute of Astrophysics, Hosakote 562114, India \\
    $^4$  Vainu Bappu Observatory, Kavalur, Alangayam 635701, India \\  
 }

\pubyear{2011}
\volume{00}
\pagerange{\pageref{firstpage}--\pageref{lastpage}}

\date{Received \today}

\maketitle
\label{firstpage}

\begin{abstract}
 An update on recent methods for automated stellar parametrization is given.
 We present preliminary results of the ongoing program for rapid
  parametrization of field stars using medium resolution spectra obtained
  using Vainu Bappu Telescope at VBO, Kavalur, India. We have used
Artificial Neural Network for estimating temperature, gravity, metallicity
and absolute magnitude of the field stars. The network for each parameter
is trained independently using a large number of calibrating stars.
The trained network is used for estimating atmospheric parameters of 
unexplored field stars.
\end{abstract}


\section{Introduction}\label{s:intro}

 The stellar spectra even at modest resolution contain  wealth of information
 on stellar parameters. In fact, most of the classification work has been
 done using medium/low resolution spectra. Hydrogen lines are good indicators
 of temperature and luminosity for a good range in spectral types; although for hotter
 end stars  lines of neutral and ionized helium, carbon and nitrogen
 are used while strengths 
 of molecular features are employed for the cool stars. A recent summary of the
  advances in  classification  can be found in Giridhar (2010).
 Additional features such as near IR triplet at 7771-74\AA~ and Ca II lines
 in 8490-8670 \AA~ region are also used for luminosity calibration.

 Many large telescopes are now equipped with
multi-object spectrometers enabling coverage of a large
number of objects per frame for stellar systems like
clusters. Instruments such as 6df on the UK Schmidt
telescope and AAOMEGA at the AAT can provide very
large number of spectra per night. On-going and future
surveys, and space missions would collect a large number
of spectra for stars belonging to different components of
our Galaxy. Such large volume of data can be handled
only with automatic procedures which would also have the
advantage of being objective and providing homogeneous
data set most suited for Galactic structure and
evolutionary studies. Another outcome would be detection
of stellar variability and finding of peculiar objects.

\section{   Automated methods for parametrization}\label{s: Automated methods}
  Several methods have been developed to estimate  atmospheric parameters from
medium-resolution stellar spectra in a fast, automatic, objective fashion.
 The most commonly adopted approaches are based upon
 the minimum distance method (MDM)  and
 those using Artificial Neural Network or ANN.
 Both the approaches use reference libraries to make comparison 
with object spectra. Other methods use
 correlations between broadband colors or the strength
of prominent metallic lines and the atmospheric parameters 
e.g. Stock and Stock (1999).

\subsection{Comparison between empirical and synthetic libraries}

 The observed stellar spectra are assigned a given spectral type and Luminosity Class (LC)
 based upon the appearance of spectral features and hence these classifications 
  are not model dependent.
 Synthetic spectra depend on model atmospheres mostly assuming
 local thermodynamic equilibrium (LTE), are affected by inadequacy of atomic
 and molecular database and non-LTE effects are severe for certain temperature/metallicity
domain. Empirical spectra however may not have the required uniform range in
 the parameter space.

 \subsection { MDM based approaches }
 The basic concept is to minimize the distance metric between the reference
 spectrum and spectrum to be classified/parametrized. The accuracy depends
 upon the density of reference spectra in parameter space.
  We need to construct a stellar spectral template library  for
 stars of known parameters. The software
 TGMET developed by  Katz et al. (1998) is based upon direct comparison
 with a reference library of stellar spectra. Soubiran et al (2003) used
this approach to estimate the T$_{\rm eff}$ , log~$g$ and [Fe/H] with
 very good accuracy  86 K, 0.28 dex and 0.16 dex respectively
 for good S/N ratio spectra of F, G and K stars. Instead of reference spectra
 synthetic spectra using the model atmospheres were used by 
 Zwitter et al.(2008) and others. In SPADES (Posbic et al. 2011) the comparison
 is made of specific lines allowing abundance determination of various elements.

\subsection { Artificial Neural Network }

A very good account of this approach can be found in numerous papers
e.g. Bailer-Jones (2002), von Hippel (1994) and others. It is a computational
method which can provide non-linear mapping between the input vector
(a spectrum for example) and one or more outputs like T$_{eff}$, log~$g$ and [M/H].
A network need to be trained with the help of spectra of stars of known parameters.
The trained network is used to parametrize the unclassified spectra.
 We have used the  back-propagation ANN code by  Ripley (1993).
 The chosen configuration of ANN is described in Giridhar, Muneer and Goswami (2006).

\section{ Analysis of VBT spectra}
 We had initiated a modest survey program for exploration of metal-poor
 candidate stars from HK Survey (Beers, Shectmann and Preston 1992), EC survey
 (Stobie et al. 1997) and high proper motion list of Lee (1984).
 The semi-empirical approach based upon the strengths of prominent lines
 and line ratio adopted in Giridhar and Goswami (2002) resulted in detection
 of several new metal-poor stars. We therefore chose to explore the use
 of ANN on a larger sample of candidate metal-poor stars.

\begin{figure}
\centerline{\includegraphics[height=9.8cm,width=12.5cm]{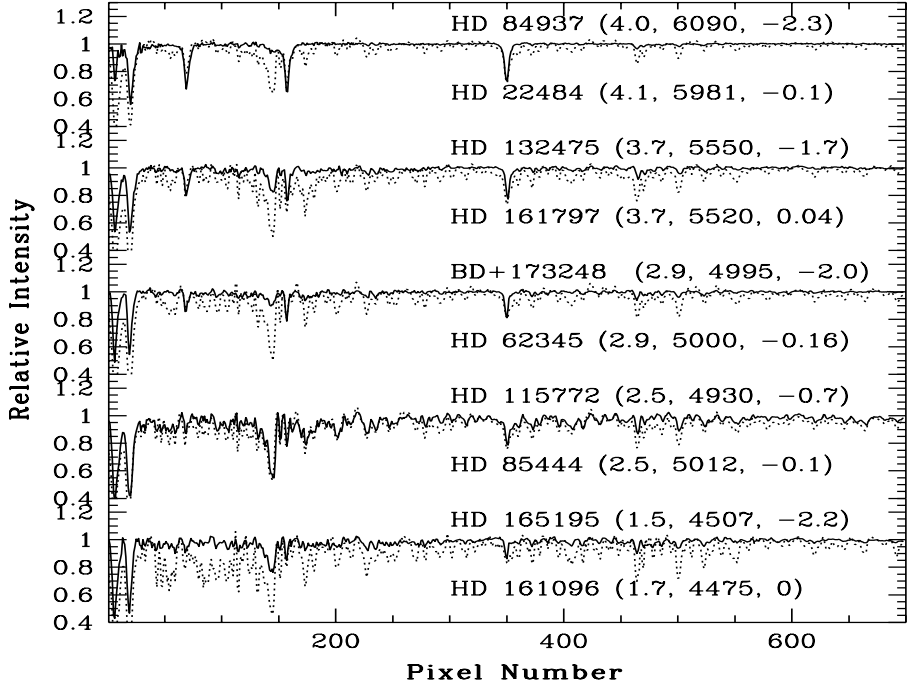}}
\caption[]{Sample spectra arranged in temperature sequence are presented.
 The stars with normal metallicity are plotted as dotted lines while metal-poor stars
 are shown as continuous lines.}
\label{sample_spectra}
\end{figure}

The medium resolution spectra (R$\sim$2000) were obtained 
using OMR spectrometer with 2.3m telescope at VBO, Kavalur.
The spectra cover 3800-6000\AA~ region. Our spectral analysis, alignment
procedure etc. are described in Giridhar, Muneer and Goswami (2006).
A few representative spectra arranged in increasing temperatures
 are presented in (Fig. \ref{sample_spectra}).

\subsection { Calibration accuracies of  stellar parameters}

Our training set containing 143 stars of known atmospheric parameters 
were chosen from Allende Prieto \& Lambert (1999), Gray, Grahm \& Hoyt(2001), 
Snider et al. (2001) and ELODIE data base (Soubiran, Karz \& Cayrel 1998).

\begin{figure} 
\centerline{\includegraphics[height=6.8cm]{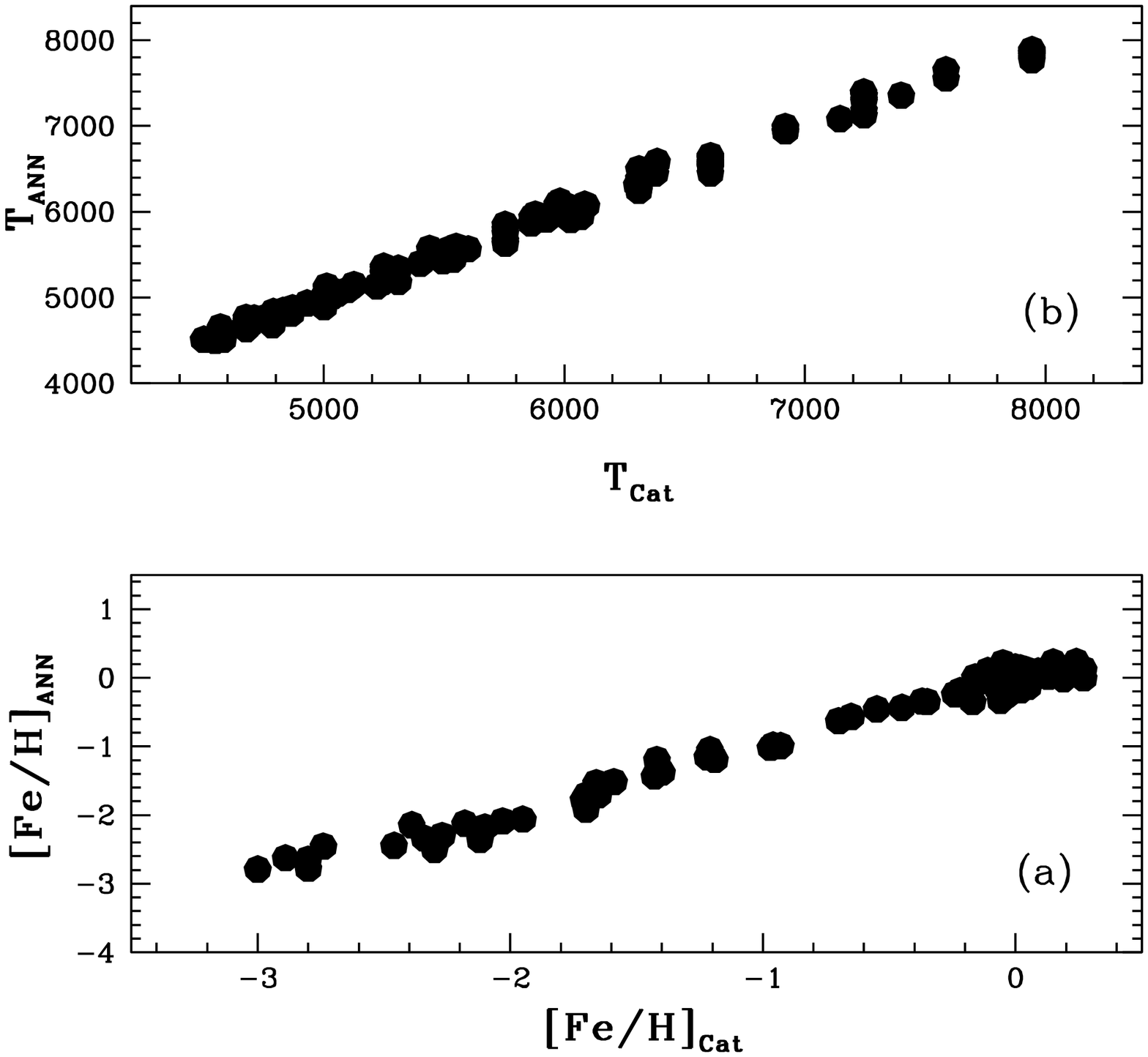} \quad 
\includegraphics[height=6.8cm]{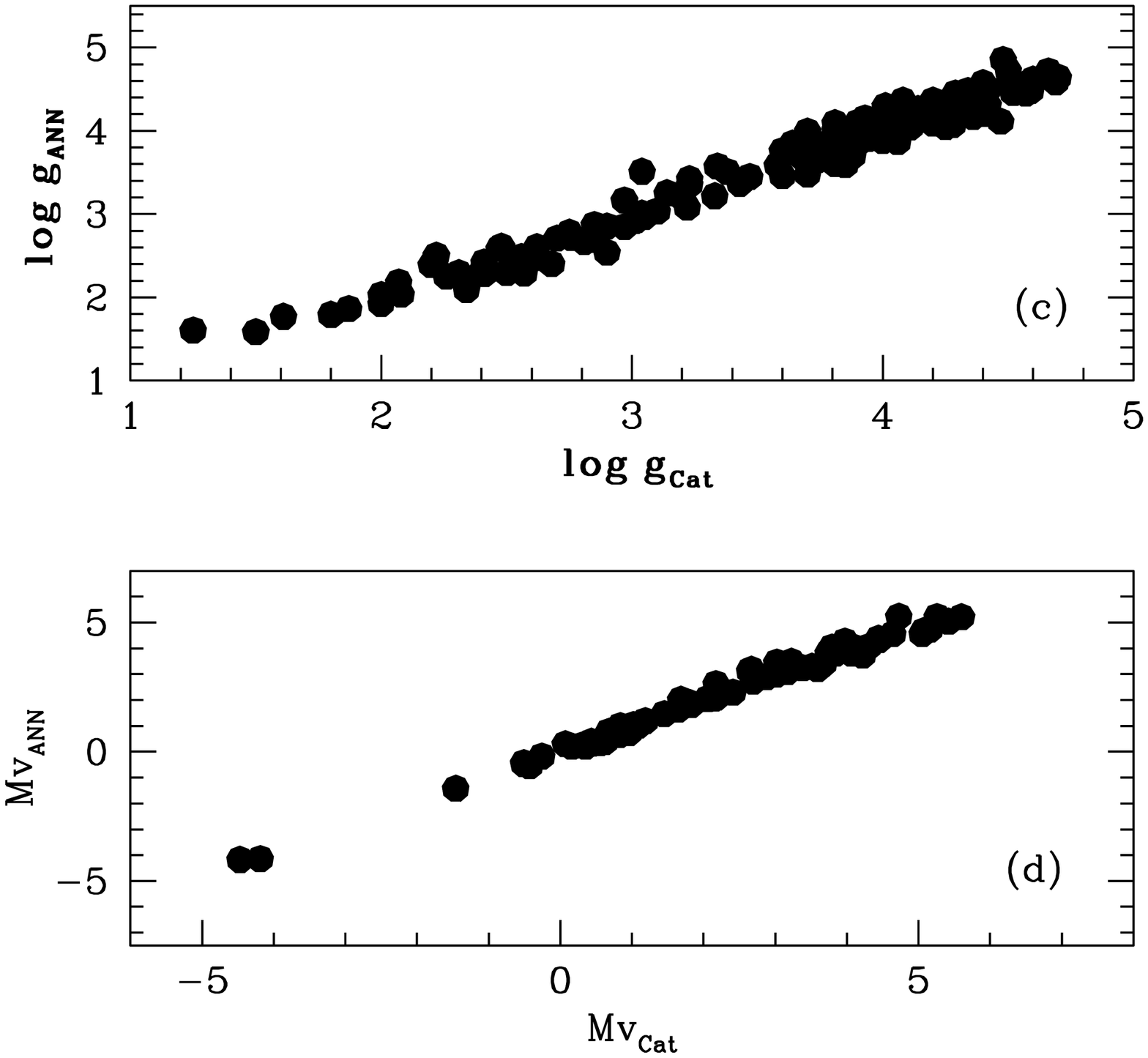}}
\caption[]{The parameters estimated from ANN are compared with those
 from literature.}
\label{Temp_metal}
\end{figure}

  Figure 2 shows the ANN results compared with calibrating values.
 We have shown in   Figure  2a [Fe/H] ANN results for 76 calibrating stars  plotted against
 those from literature. For the metallicity range of $-$3.0 to $+$0.3 dex. the RMS scatter
 about the line of unity is 0.3 dex which  is  similar to the intrinsic uncertainties
 metallicities for calibrating stars.
 To avoid using the same spectra for training and testing
 purposes, we divided the training set into two parts and trained ANN for each part.
 Then the weights for part~1 were used to estimate [Fe/H] for stars in part~2 while
 those of part~2 were used to estimate [Fe/H] for stars in part ~1. The errors
 shown in the Figure 2 are therefore realistic estimate of errors.
 This approach of dividing calibrating sample into two separate training
 and testing sets has been adopted for T$_{\rm eff}$ and log~$g$  calibration also.

  We had good T$_{\rm eff}$ and log~$g$ estimates for 143 stars for calibrations 
and among them 110 stars had nearly solar metallicities while 33 were hard core metal-poor
 stars. While training the networks for temperature we found that usage of the same ANN 
 for normal metallicity stars as well as metal-poor stars was giving large calibration errors
 (250 to 300K for T$_{\rm eff}$).
 It is understandable as the spectra of metal-poor stars and also those of hot stars have
 weak metallic lines. To overcome this degeneracy 
 we used separate ANNs for each metallicity subgroup for the temperature (as well as gravity)
 calibration. The temperatures estimated by AAN are
 compared with the literature values in Figure 2b. The RMS error is now reduced to 150K.
 
 The Figure 2c  shows the result for gravity calibration adopting the
  procedure mentioned above. 
 The RMS error is about 0.35 for log ~$g$ range of 1 to $+$4.5 dex.

 A large fraction of stars observed by us have good parallax estimates (errors less than 20\%).
 Combining the V magnitudes with parallaxes the distances and hence M$_{V}$ could be 
 estimated. Most of these objects were nearby objects so the effect of
 interstellar extinction could be assumed as negligible.
 Our spectral region contains many luminosity sensitive features like hydrogen lines,
 Mg I lines at 5172-83\AA~, G bands etc. 
 However, the same feature cannot serve the whole range of spectral types. We have divided 
 the sample stars into two temperature groups and yet another group for metal-poor objects.
 The usage of three separate networks helped in attaining calibration accuracy
 $\sim$  $\pm$0.3 mag for M$_{V}$. The  M$_{V}$ estimated by ANN are compared with those 
 estimated from parallaxes  as shown in Figure 2d.

\section{ Stellar parameters for metal-poor candidate stars}

 A different set of 
  ANNs  for each  atmospheric parameter were trained for  metal-poor candidate stars.
 A preliminary estimation of metallcity was made using ANN trained  on the full range of
 metallicity. Then, we refined the measurements by using  two different ANN sets; one for   
  estimating the atmospheric parameters for stars of near solar metallicties and the other for
 the significantly metal-poor stars ([Fe/H] $<$ $-$0.7 dex).  The (B-V) colours were available for
 many of them which were used to verify the T$_{\rm eff}$ estimated by ANN. In most cases
 the temperature estimated using ANN were in close agreement with colour temperatures.
 A sizeable fraction of the candidate stars belonged to [Fe/H] $-$0.5 to $-$2.5 dex range.

\section{ Conclusions}

We have demonstrated that using ANN we can measure atmospheric parameters
with an accuracy of $\pm$ 0.3dex in [Fe/H], $\pm$200K in temperature and $\pm$0.35 in log~$g$
with the help of training set of stars of known parameters. We find that independent 
calibrations for near solar metallicity stars and metal-poor stars decrease the 
errors in T$_{\rm eff}$ and log~$g$ by a factor of two. We have extended the application of this
 method to estimation of absolute magnitude using nearby stars with well determined parallaxes.
 Better M$_{V}$ calibration accuracy can be obtained by using two separate ANNs for cool
 and warm stars. The present accuracy of  M$_{V}$ calibration is $\sim$$\pm$0.3mag.   

\section {Acknowledgment}

This work was partially funded by the National Science 
Foundation's Office of International Science and 
Education, Grant Number 0554111: International Research 
Experience for Students, and managed by the National Solar 
Observatory's Global Oscillation Network Group.

\label{lastpage}
\end{document}